\documentclass[journal]{vgtc}           

\onlineid{1123}

\vgtccategory{Application or Design Study}

\title{
  \texorpdfstring
  {\texttt{lpviz}: Interactive Linear Programming Visualization}
  {lpviz: Interactive Linear Programming Visualization}
}
\author{%
  \authororcid{Evan Grand}{0009-0003-6237-8092} and
  \authororcid{Michael Klamkin}{0000-0002-2131-8603}
}

\authorfooter{
  \item
  	Evan Grand is with New York University.
  	E-mail: egrand@nyu.edu.
  \item
  	Michael Klamkin is with the Georgia Institute of Technology.
}

\abstract{%
  This paper presents \lpviz{}, a browser-based visualization tool for linear programming. \lpviz{} is deeply interactive, offering an intuitive interface where users can directly draw and edit the feasible region and objective vector, without requiring cumbersome manipulation of raw numerical coefficients. \lpviz{} lets users compare the behavior of several classes of linear programming algorithms, namely Simplex, Interior-Point, Primal-Dual Hybrid Gradient, and Central Path. In the 3D mode, \lpviz{} places iterates at heights corresponding to important solver metadata such as complementarity gap or KKT residual, helping users gain further insight into algorithm behavior beyond the primal iterates alone. \lpviz{} has been used in both research and classroom settings, to help develop intuition for the strengths and weaknesses of different solvers and the impact of solver settings on convergence behavior. \lpviz{} is open-source, permissively licensed, and freely available on any device with a web browser at \url{https://lpviz.net}.
}


\teaser{
  \centering
  \begin{subfigure}[b]{0.32\columnwidth}
  	\centering
  	\includegraphics[width=\textwidth, alt={Screenshot of lpviz showing IPM trajectories in Trace mode.}]{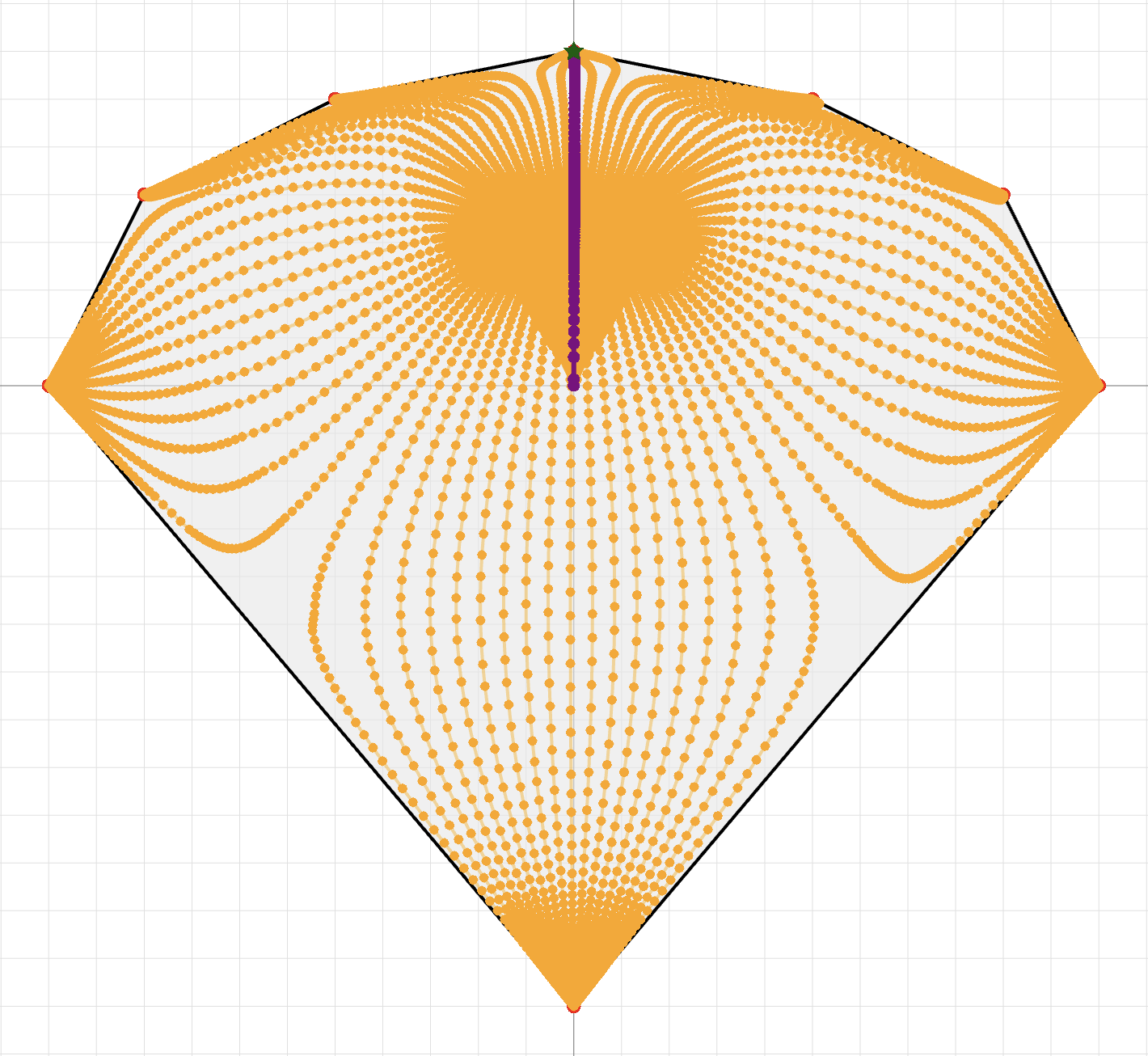}
  	\caption{Interior Point Method}
  	\label{fig:teaser:ipm}
  \end{subfigure}%
  \hfill%
  \begin{subfigure}[b]{0.32\columnwidth}
  	\centering
  	\includegraphics[width=\textwidth, alt={Screenshot of lpviz showing the central path in Trace mode.}]{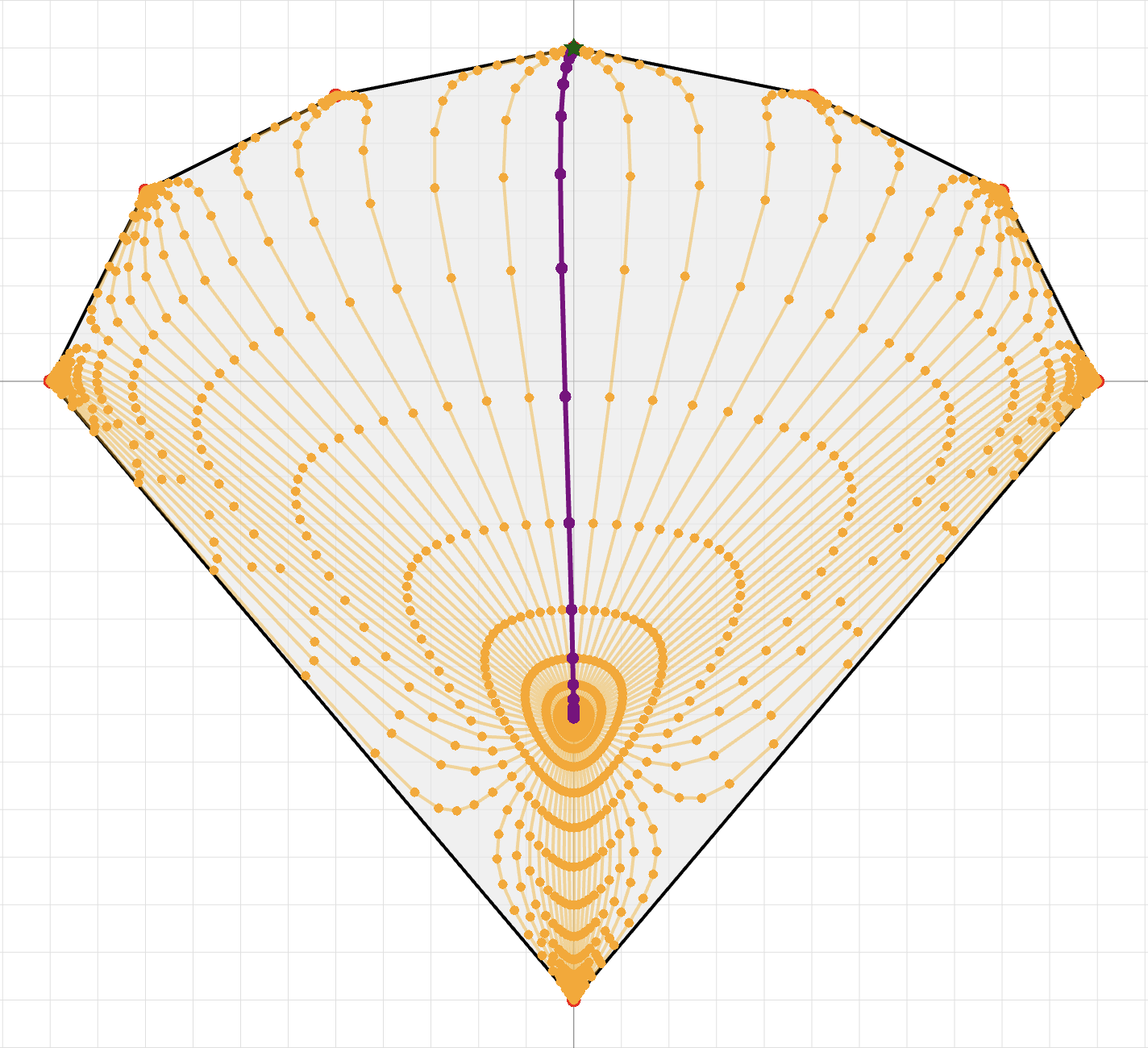}
  	\caption{Central Path}
  	\label{fig:teaser:cp}
  \end{subfigure}%
  \hfill
\begin{subfigure}[b]{0.32\columnwidth}
  	\centering
  	\includegraphics[width=\textwidth, alt={Screenshot of lpviz showing PDHG iterates together with restarted Halpern behavior.}]{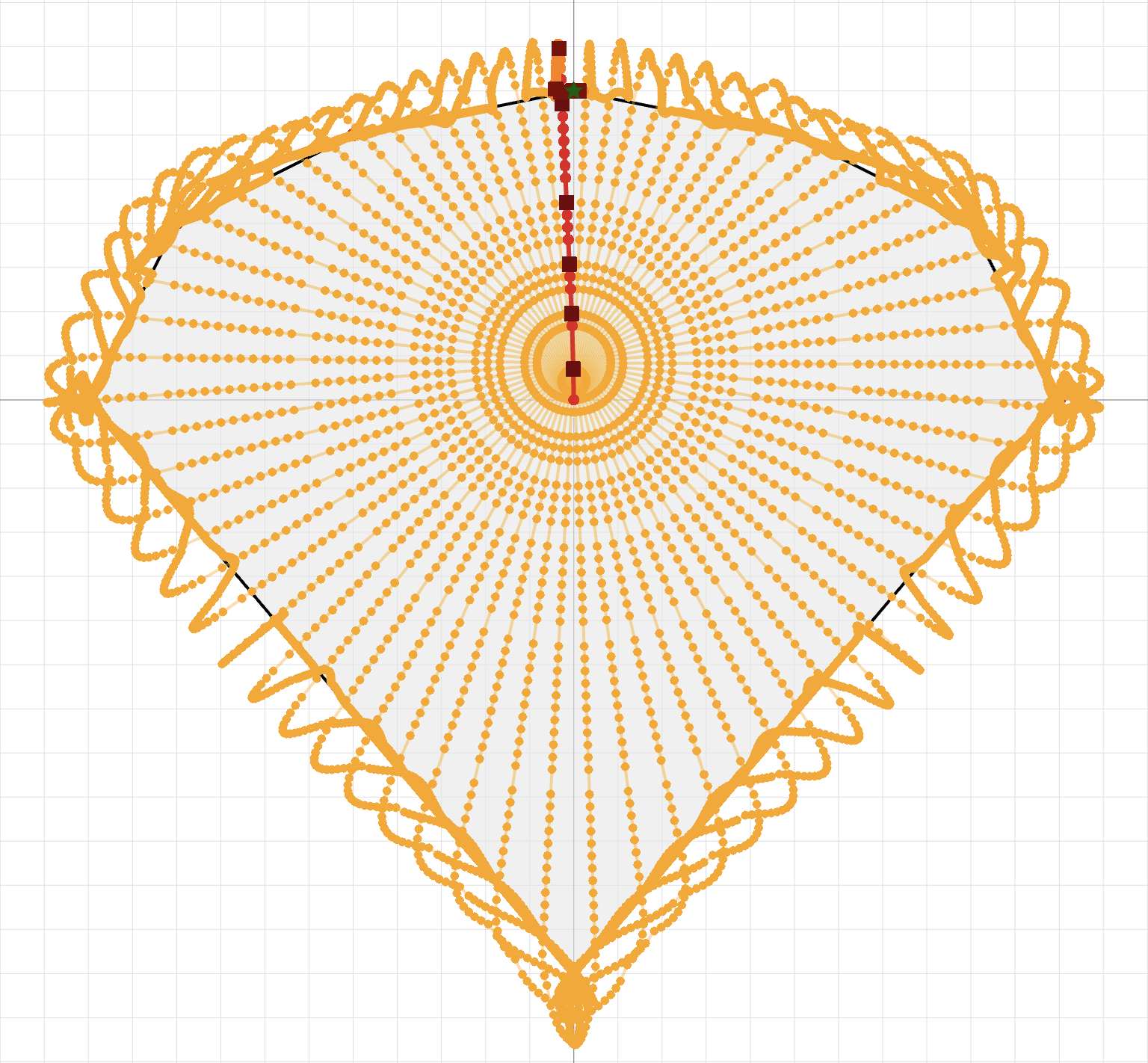}
  	\caption{PDHG with Restarted Halpern}
  	\label{fig:teaser:pdhg}
  \end{subfigure}%
  \\
  \begin{subfigure}[b]{\columnwidth}
    \centering
  	\includegraphics[width=0.9\textwidth, alt={Screenshot of lpviz in 3D mode, showing the feasible region and IPM trajectory under objective rotation.}]{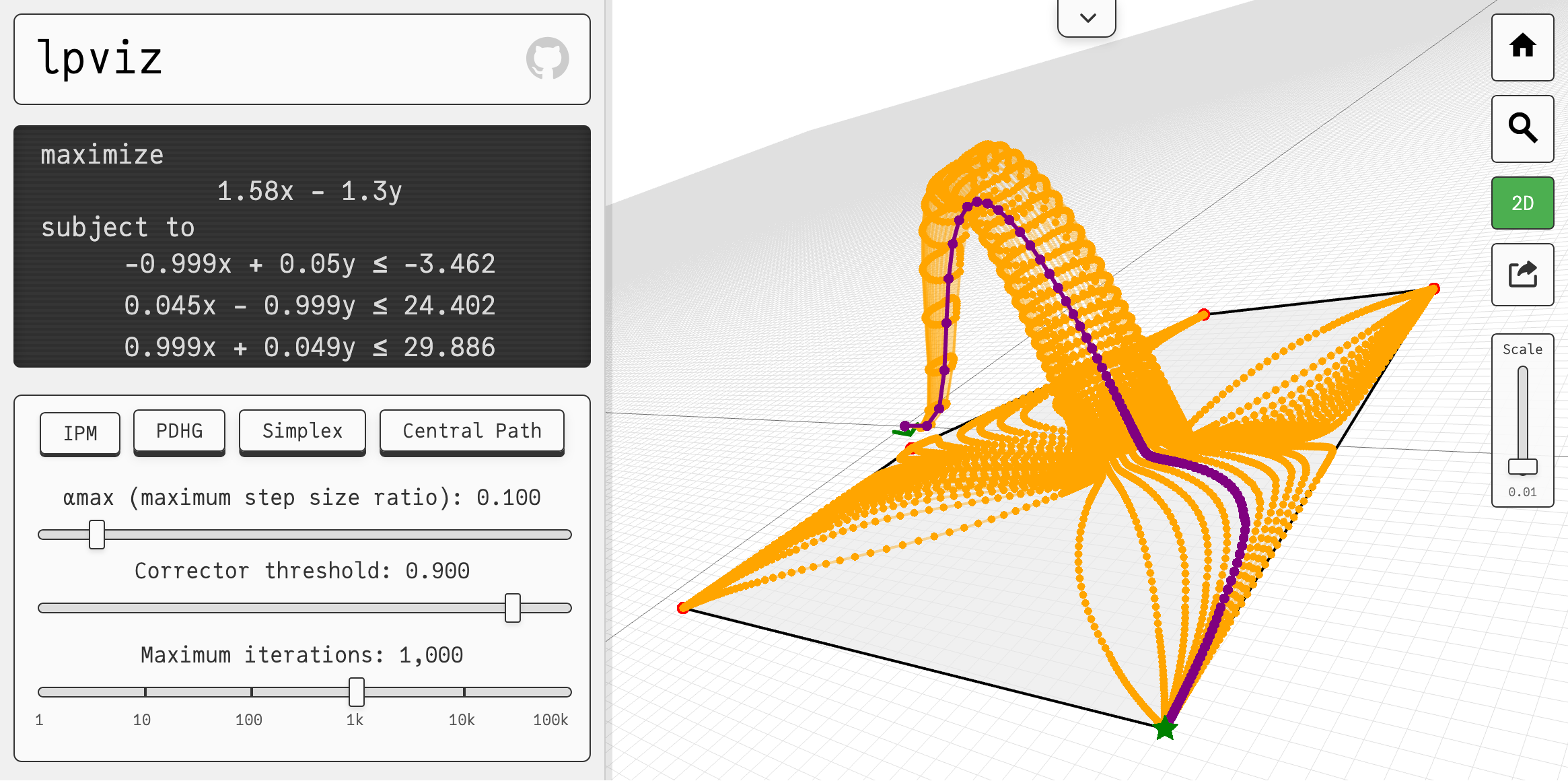}
  	\caption{Interior Point Method in 3D mode}
    \label{fig:teaser:3d}
  \end{subfigure}
  \subfigsCaption{Screenshots of \lpviz{} in \textcolor{orange}{\texttt{Trace}} mode, visualizing how convergence behavior changes as the objective rotates.}
  \label{fig:teaser}
}

\renewcommand{\manuscriptnotetxt}{}

\graphicspath{{figs/}{figures/}{pictures/}{images/}{./}} 

\usepackage{tabu}                      
\usepackage{booktabs}                  
\usepackage{lipsum}                    
\usepackage{mwe}                       
\usepackage{ccicons}                   

\usepackage{mathptmx}                  

\usepackage{amsmath,amssymb,mathtools}
\usepackage{algorithm}
\usepackage{algorithmic}

\newcommand{\lpviz}{\texttt{lpviz}}

\begin{document}
\firstsection{Introduction}
\maketitle

    Linear programming (LP) is a core building block in engineering computation,
    finding many applications in diverse fields including
    power grid operations, supply chain planning,
    sports scheduling, manufacturing, finance, and more.
    Despite the importance and ubiquity of linear programming, the inner workings of
    LP solver algorithms remain a mystery to many students and practitioners.
    Indeed, a key feature of mathematical programming is to transform
    the problem-solving task from
    ``how to write a custom algorithm to solve the problem'' to
    ``how to model the problem such that a generic algorithm can solve it.''
    Thus, undergraduate courses covering linear programming
    tend to focus more on modeling techniques than on
    solver algorithm details. However, it is important to develop intuition
    for how different families of solver algorithms behave in different circumstances,
    in order to know when to use which one, as well as to better anticipate and diagnose issues.
    Furthermore, recent years have seen a resurgence in linear programming solver developments spurred by advances in GPU hardware \cite{lu2024restarted}, whose behavior can differ substantially from classical methods.
    
    To help students, practitioners, and researchers develop intuition for how different LP solver algorithms behave,
    this paper presents \lpviz{}, a browser-based visualization tool for LP. By leveraging modern web development techniques, \lpviz{} offers a responsive and deeply interactive interface, allowing users to click and drag to specify and modify the linear program rather than typing numerical coefficients manually. Freely available at \url{https://lpviz.net}, it requires no setup or sign up, and can run on any desktop or mobile device with a large enough screen.

\subsection{Background}

    To allow for plotting in 2D, \lpviz{} considers two-variable linear programs, i.e. optimization problems that can be written as
    \begin{align*}
       \text{(LP)} \quad \max_{x_1,x_2} \quad & c_1 x_1 + c_2 x_2 \\
        \text{s.t.} \quad & a^{(j)}_1 x_1 + a^{(j)}_2 x_2 \leq b^{(j)}\quad\forall j \in [1,m]
    \end{align*}
    where $c_1,\,c_2\in\mathbb{R}$ are the objective vector coefficients, $a^{(j)}_1,\,a^{(j)}_2\in\mathbb{R}$ are the $j^\text{th}$ constraint coefficients, $b^{(j)}$ is the $j^\text{th}$ right-hand-side, and $m$ is the number of constraints. In \lpviz{} the objective vector $\begin{bmatrix}
        c_1\enspace c_2
    \end{bmatrix}^\top$ is visualized with a green arrow and the feasible region \mbox{$\mathcal{X}\coloneq\{x\mid a^{(j)}_1 x_1 + a^{(j)}_2 x_2 \leq b_j \quad\forall j \in [1,m]\}$} is visualized using a solid shaded area with $m$ edges, each corresponding to a constraint $j$.

    \subsection{Related Work}
This section briefly reviews select recent work on Algorithm Visualization (AV) \cite{shaffer2010algorithm} for optimization algorithms.
            GILP \cite{robbins2023gilp} and gMOIP \cite{nielsen2020gmoip} are Python-based visualizations of the Simplex algorithm. Unlike \lpviz{}, GILP and gMOIP support 3D problems, and gMOIP supports multi-objective optimization and integer variables. Instead, \lpviz{} focuses on the canonical LP form; single-objective and continuous variables, and supports only 2D problems so that the Z-axis can be used to visualize algorithm metadata.
            Another line of work considers how to visualize higher-dimensional LPs using projections \cite{espadoto2021optmap,olkhovsky2022visualizing}. Though valuable for e.g. human-in-the-loop approaches \cite{cajot2019interactive}, \lpviz{} intentionally supports only 2D problems, avoiding artifacts related to projection and removing the need for manual numerical input. 
            AV has also been applied to optimization outside of LP, including visualizing the behavior of deep learning optimizers\footnote{\url{https://github.com/jettify/pytorch-optimizer?tab=readme-ov-file\#visualizations}} and genetic algorithms \cite{genetic}.

\subsection{Contributions}

The core contributions of this paper are:
\begin{itemize}
    \item It presents \lpviz{}, a browser-based interactive visualization tool for linear programming that runs entirely client-side, requiring no installation or server-side computation beyond file hosting.
    \item It proposes an intuitive interface for specifying and modifying two-variable linear programs, allowing users to interactively construct and edit the feasible region and objective direction, watching the solver iterates update in real-time. This removes the need for cumbersome manual input of numerical coefficients.
    \item \lpviz{} offers a unified environment for the visualization of four families of linear programming methods -- Simplex, Interior Point, Primal-Dual Hybrid Gradient, and Central Path. This enables side-by-side exploration of how solver trajectories differ on the same problem or family of problems. 
    \item It describes how modern browser technologies such as Web Workers, \texttt{TypedArray}-based numerical kernels, and WebGL-based GPU-accelerated rendering are combined to implement real-time in-browser optimization and responsive rendering.
\end{itemize}

\section{System Description}

This section describes the internal design of \lpviz{}. A key feature of \lpviz{} is that it runs fully on-device: all computations run entirely via static web assets in the browser. This design removes network latency while allowing \lpviz{} to be accessible without installation. Many Content Delivery Networks (CDNs) also host static files for free, reducing hosting costs.\footnote{\lpviz{} uses Cloudflare, which offers free hosting through Workers.}

\begin{figure}[t]
    \centering
    \includegraphics[width=0.95\columnwidth, alt={System architecture diagram for lpviz.}]{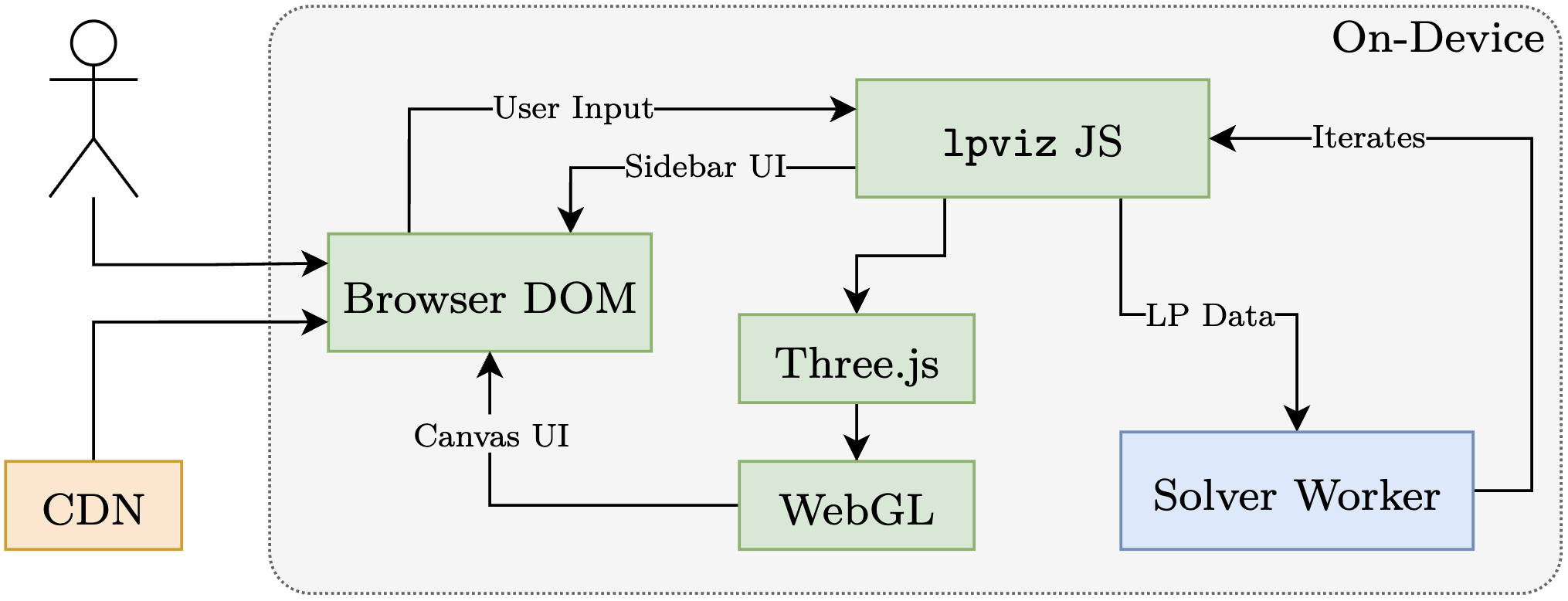}
    \caption{System architecture diagram for \lpviz{}.}
    \label{fig:architecture}
\end{figure}

The overall architecture is visualized in Figure \ref{fig:architecture}. The Browser downloads \lpviz{} (HTML, JavaScript (JS), and CSS files) from the CDN, then waits for user inputs. Interaction is handled by the \lpviz{} JS module, which forwards the problem data (objective and constraint coefficients) to the Solver Worker, and emits rendering instructions to WebGL via the Three.js interface. Although the solvers are implemented using TypeScript (thus in principle could be included in the \lpviz{} JS module), as discussed in Section \ref{sec:perf}, this split architecture allows the solver computation and rendering workloads to run on separate threads, ensuring smooth canvas updates while the solver is running.

\subsection{Interactive Problem Definition}

Unlike previous AV for optimization tools, \lpviz{} offers a unique and intuitive problem definition interface. Rather than requiring manual input of raw objective and constraint function coefficients, users simply click on the canvas grid to incrementally build the feasible region. It is important to note that this interface corresponds to defining the feasible region as the convex hull of its vertices, i.e. the ``V-representation'', rather than in terms of the canonical ``H-representation'' which defines the feasible region as the intersection of a set of half-spaces. In practice, the ``V-representation'' is unpopular since the vertices are not known ahead of time in general. However, since \lpviz{} considers only two-dimensional LPs, it is computationally tractable to gradually convert user input vertices to half-spaces, which can then be passed to the solvers. Let $v_i$ denote the $i^\text{th}$ vertex input by the user. On each vertex input after the first, \lpviz{} computes the line connecting $v_{i+1}$ to $v_i$. This approach guarantees that all $v_i$ become vertices of the final feasible region since \lpviz{} rejects vertex inputs that would otherwise result in nonconvex regions.

\begin{figure}[h!]
  \centering
  \begin{subfigure}[b]{0.48\columnwidth}
  	\centering
  	\includegraphics[width=\textwidth, alt={Spiral behavior of PDHG in Inequality mode on an unbounded problem.}]{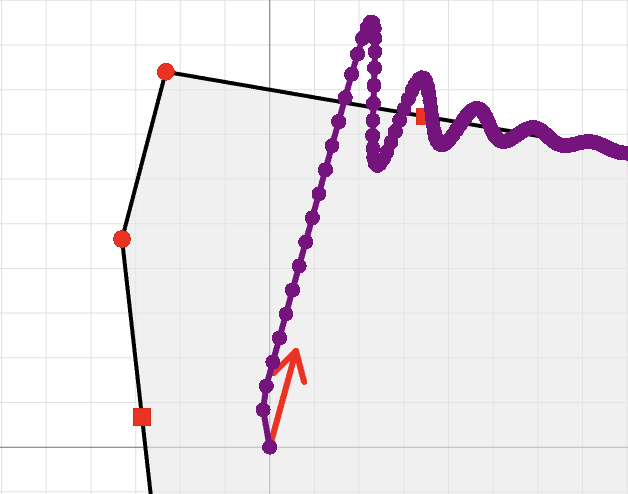}
  	\caption{PDHG in Inequality mode}
  	\label{fig:unbounded:ipm}
  \end{subfigure}%
  \hfill%
  \begin{subfigure}[b]{0.48\columnwidth}
  	\centering
  	\includegraphics[width=\textwidth, alt={Convergence behavior of IPM on an unbounded problem.}]{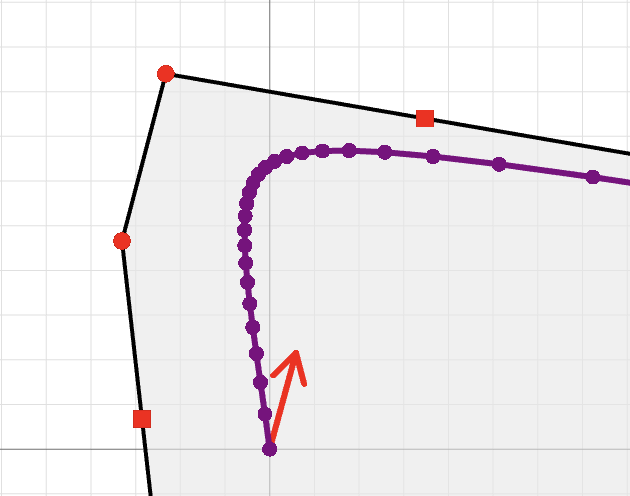}
  	\caption{Interior Point Method}
  	\label{fig:unbounded:pdhg}
  \end{subfigure}%
  \caption{Comparison of IPM and PDHG on an unbounded problem.}
  \label{fig:unbounded}
\end{figure}

To allow users to model unbounded feasible regions, \lpviz{} offers two drawing termination signals -- clicking inside the region or close to the first vertex creates a bounded region, while pressing enter creates an unbounded region with the first and last lines extending to infinity. Unlike the bounded-region case where all objective directions are valid, in the unbounded-region case it is possible set an objective direction that renders the problem unbounded (i.e., no optimal solution exists). \lpviz{} colors the objective arrow red to signal unboundedness, but still allows the user to run the solvers since each algorithm handles unbounded problems differently; this is visualized in Figure \ref{fig:unbounded}.

\subsection{Solving Linear Programs in the Browser}\label{sec:perf}

\lpviz{} encourages the user to interactively adjust the objective direction, feasible region, and solver settings, to see how the different solver algorithms react to small changes. For instance, users can drag vertices of the feasible region, use the Rotate Objective button, or even delete vertices, and see the solver iterates update smoothly on the canvas in real-time. To enable such responsive updates, \lpviz{} includes high performance implementations of the underlying solver algorithms. To implement LP solvers efficiently in TypeScript, \lpviz{} includes custom kernels for key subroutines implemented using the \texttt{TypedArray} abstraction. These include an LU factorization routine for solving linear systems, a matrix-vector product kernel, and other basic linear algebra operations. \texttt{TypedArray} offers an array-like view of a contiguous binary buffer, meaning the browser's JavaScript engine can generate efficient specialized code for these core numerical operations. Finally, to ensure the user interface does not freeze while a solver is running, solvers run on a dedicated thread using HTML Web Workers. Table \ref{tab:solverperf} compares equivalent implementations differing only in the array type and allocation pattern; pre-allocation and \texttt{TypedArray} give a clear speedup over the naive eager \texttt{number[]} approach\footnote{The na\"{\i}ve approach corresponds to an earlier version of \lpviz{} which used the \texttt{ml-matrix} library rather than custom \texttt{TypedArray} kernels; experiments were run on an Apple M4 Max CPU.}.

\begin{table}[h]
\centering
\caption{Performance comparison of na\"{\i}ve and optimized solver implementations, on a problem with 20 constraints.}
\label{tab:solverperf}
\begin{tabular}{l|rr|r}
\toprule
Solver & Optimized (ms) & Na\"{\i}ve (ms) & Speedup \\
\midrule
Central Path   & 0.61  & 4.17  & 6.81$\times$ \\
IPM            & 0.79 &  4.16  & 5.29$\times$ \\
Simplex        & 0.74  & 1.78  & 2.40$\times$ \\
PDHG (eq)      & 5.21  & 95.52 & 18.32$\times$ \\
PDHG (ineq)    & 0.44  & 6.61  & 15.13$\times$ \\
\bottomrule
\end{tabular}
\end{table}
{
Besides solving LPs, the main computational workload of \lpviz{} is (re-)rendering the canvas. To improve performance of the render loop, the canvas is decomposed into a stack of Three.js \texttt{Scene} layers, which are each updated independently and only when needed. This allows \lpviz{} to minimize rendering latency by re-using work from previous iterations when geometry stays the same. This is particularly important during 3D navigation: moving the camera changes the view of the scene but does not change the grid, polytope, objective, trace, or iterate geometries. Thus, camera-only updates can be rendered without recomputing every layer. The most computational intensive rendering setting in \lpviz{} is traced objective rotation, which creates a large number line segments and points. To address this, \lpviz{} batches trace geometries into one mesh, avoiding re-rendering thousands of Three.js \texttt{Line2} objects each frame.

Table~\ref{tab:renderperf} shows results from an ablation study using a stress-test case where PDHG is used with a very small step size (0.001) and a large maximum iteration count (10,000); when combined with a traced objective rotation, this creates millions of points and line segments. Two workloads are measured: traced objective rotation (i.e., solving and rendering) and a 3D camera movement (i.e., only rendering). The results show that batching trace lines is important for both workloads, reducing latency by \(9.3\times\) during traced rotation and \(12.5\times\) during 3D camera movement. Fine-grained geometry invalidation, though it increases the implementation complexity, has an extreme impact on performance: avoiding unnecessary updates reduces latency by \(99\times\) during traced rotation and \(250\times\) during 3D camera movement. Together, the solver and rendering performance improvements let \lpviz{} run in real-time, responding instantly to user input in most settings.

\begin{table}[t]
\centering
\caption{Rendering performance ablation for a traced quarter-rotation using PDHG equality on a problem with 4 constraints,
 \texttt{maxit}=1000, and angle step \(0.001\). This produces 1571 traces, amounting to \textasciitilde1M points.}
\label{tab:renderperf}
\begin{tabular}{l@{\hspace{0em}}rrrr}
\toprule
& \multicolumn{2}{c}{Traced Rotate} & \multicolumn{2}{c}{3D Camera Move} \\
\cmidrule(lr){2-3}\cmidrule(lr){4-5}
Variant & Time (ms) & vs Opt. & Time (ms) & vs Opt. \\
\midrule
{Optimized} & {0.3} & {1.0$\times$} & {0.2} & {1.0$\times$} \\
\midrule
Unbatched trace geometry & 2.8 & 9.3$\times$ & 2.5 & 12.5$\times$ \\
All-layer updates & 30 & 99$\times$ & 50 & 250$\times$ \\
\bottomrule
\end{tabular}
\end{table}
}

\section{Solver Algorithms}

\begin{figure}[h]
        \centering
        \includegraphics[width=0.95\columnwidth, alt={An example where the difference-of-positives reformulation introduces additional phantom simplex pivots at axis intersections.}]{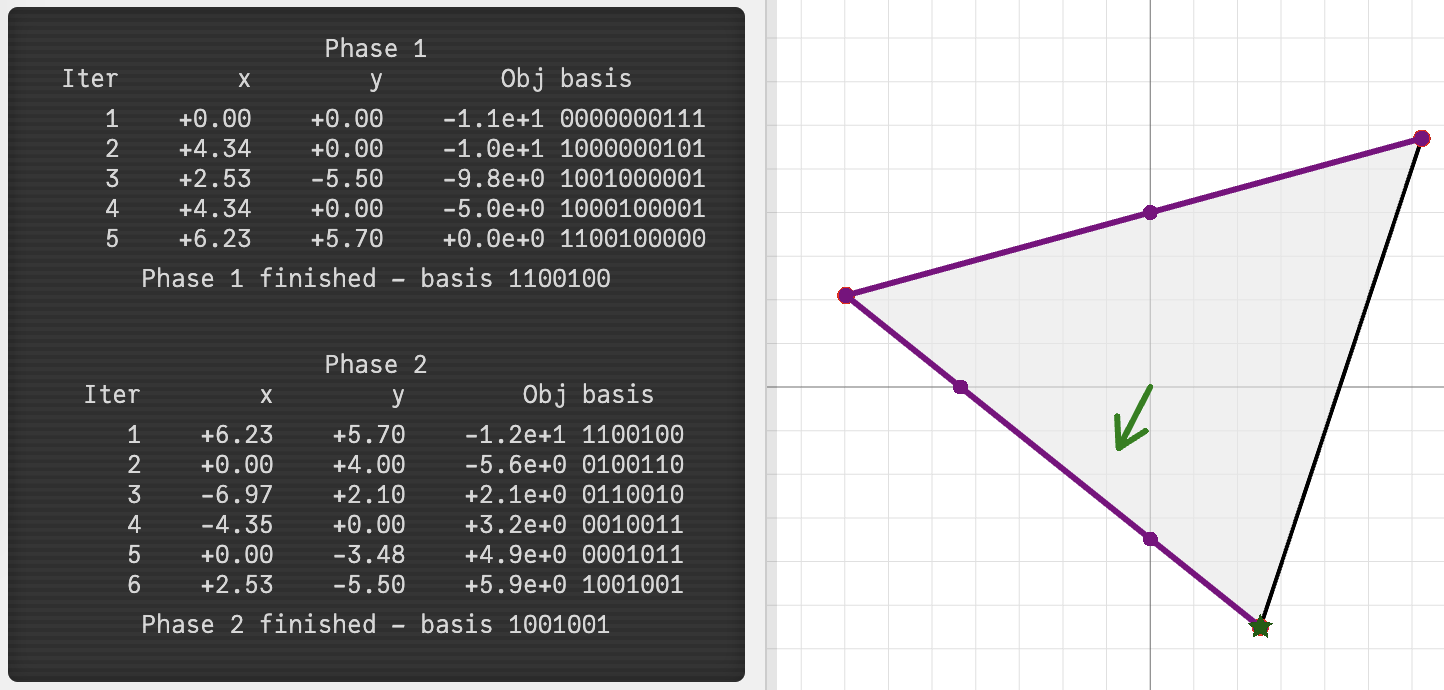}
        \caption{Phantom pivots caused by difference-of-positives reformulation.}
        \label{fig:simplexdop}
    \end{figure}

\begin{figure}[b!]
  \centering
  \begin{subfigure}[b]{0.45\columnwidth}
  	\centering
  	\includegraphics[width=0.9\textwidth, alt={PDHG in Equality mode.}]{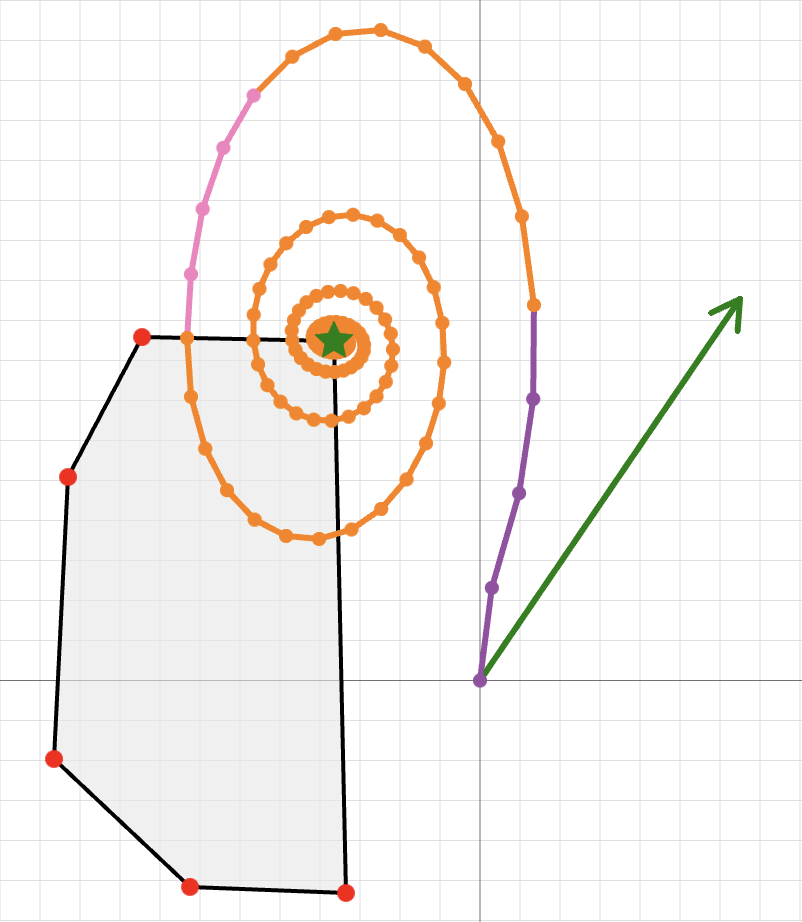}
  	\caption{PDHG in Inequality mode.}
  	\label{fig:eqineq:ineq}
  \end{subfigure}%
  \hfill%
  \begin{subfigure}[b]{0.45\columnwidth}
  	\centering
  	\includegraphics[width=0.9\textwidth, alt={PDHG in Inequality mode (on the same problem as Equality mode).}]{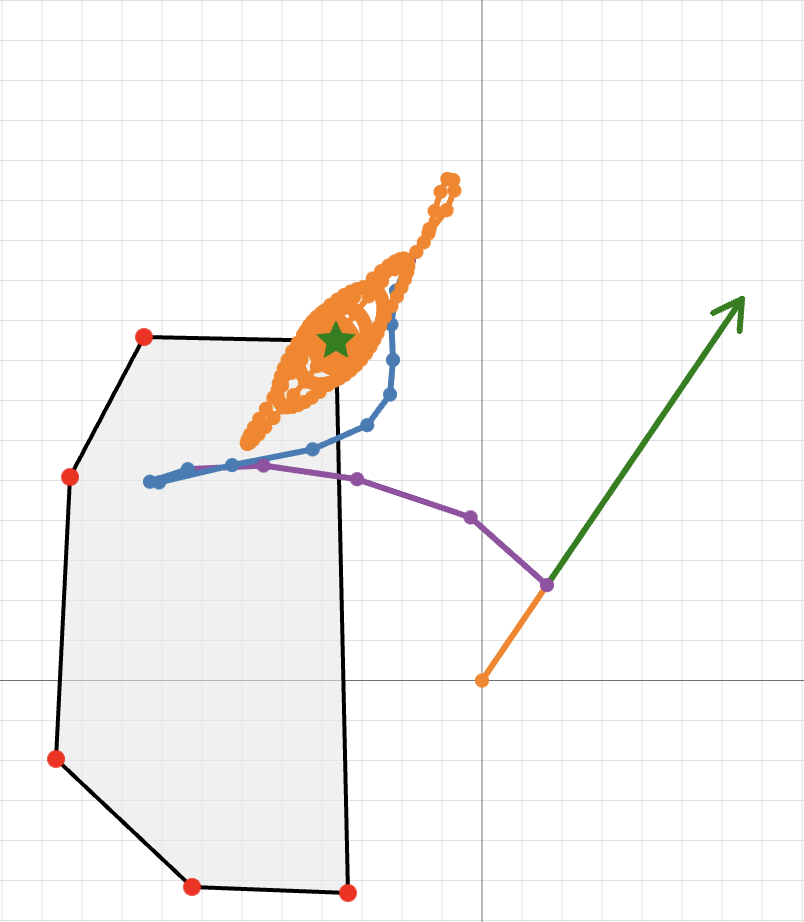}
  	\caption{PDHG in Equality mode.}
  	\label{fig:eqineq:eq}
  \end{subfigure}%
  \caption{Comparison of Inequality and Equality modes in PDHG.}
  \label{fig:eqineq}
\end{figure}

This section describes the four families of LP solver algorithms included in \lpviz{}. To stay close to textbook descriptions, each algorithm uses its own standard form; for instance the Simplex algorithm requires positive variables and equality constraints, while the original problem in \lpviz{} uses inequality constraints and unconstrained variables. This reformulation is done prior to passing the problem to each solver, making sure to convert the output data to the original space for plotting. Figures \ref{fig:simplexdop} and \ref{fig:eqineq} show the impact of these reformulations.

\paragraph{Simplex}
    The simplex method moves along edges of the feasible region, visiting
    adjacent vertices until no improving move remains. \lpviz{} implements a
    two-phase revised simplex method \cite{wagner1957comparison}.
    The algorithm chooses the first nonbasic column with positive reduced cost to enter the basis, and
    breaks ratio-test ties using Bland's rule \cite{bland1977new}. Since the algorithm requires positive variables and equality constraints, the implementation
    first converts variables to the difference-of-positives form $x=x^+-x^-$ where $x^+,x^-\geq 0$ and adds slack variables. The solver relies on a shared core pivoting routine; the first call to this routine solves the so-called Phase I problem to determine an initial feasible basis, and the second call starts from this basis to solve the original LP. Note that \lpviz{} only visualizes the Phase II iterates on the canvas, and that the drawing logic guarantees a feasible basis exists.

    Due to the difference-of-positives reformulation, not all pivots correspond to vertices of the original feasible region. In particular,
    additional pivots appear where the original edges intersect the coordinate axes (corresponding to the sign constraints of $x^+$ and $x^-$). This is visualized in Figure \ref{fig:simplexdop} where Simplex uses 6 Phase II iterations on a problem with only 3 original vertices.

\begin{figure}[h!]
  \centering
  \begin{subfigure}[b]{0.45\columnwidth}
  	\centering
  	\includegraphics[width=0.9\textwidth, alt={PDHG in Equality mode.}]{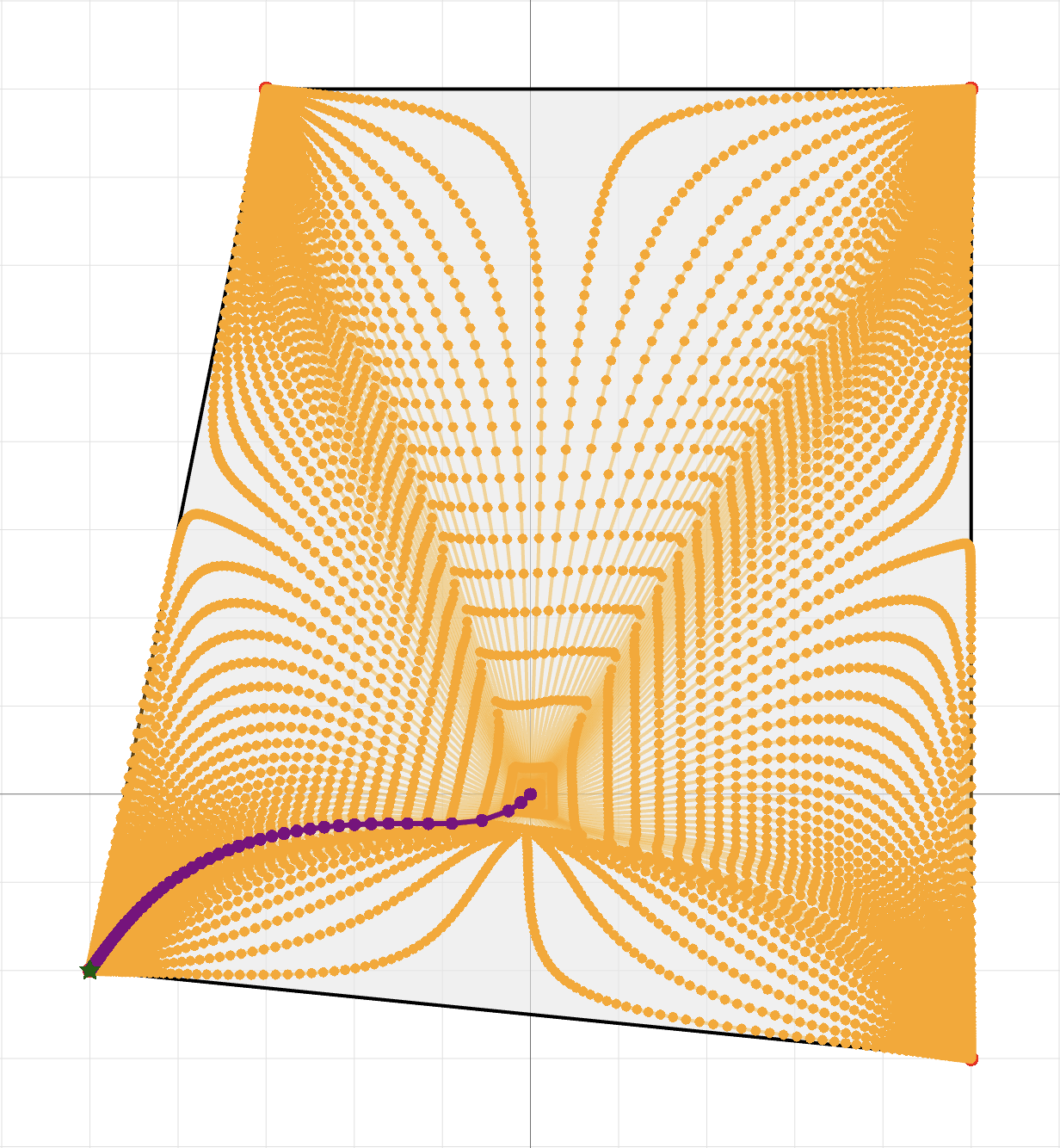}
  	\caption{High corrector threshold.}
  	\label{fig:corr:high}
  \end{subfigure}%
  \hfill%
  \begin{subfigure}[b]{0.45\columnwidth}
  	\centering
  	\includegraphics[width=0.9\textwidth, alt={PDHG in Inequality mode (on the same problem as Equality mode).}]{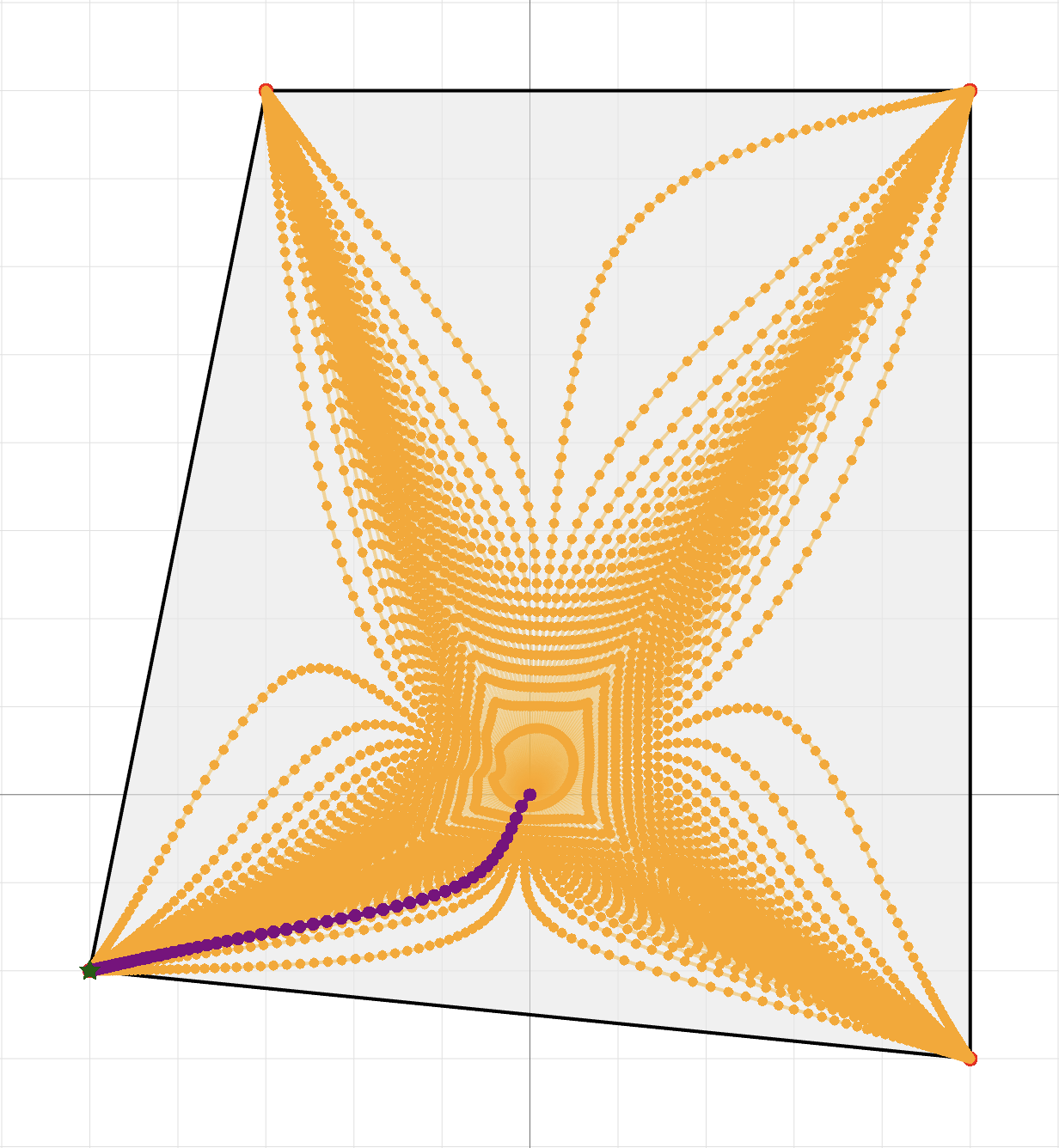}
  	\caption{Low corrector threshold.}
  	\label{fig:corr:low}
  \end{subfigure}%
  \caption{Comparison of IPM with high and low corrector thresholds.}
  \label{fig:corr}
\end{figure}

\paragraph{Interior-Point Method}
    There are several variants of Interior-Point Methods (IPM); \lpviz{} implements an
    infeasible primal-dual predictor-corrector method \cite{mehrotra1992implementation}.
    The algorithm starts
    from the initialization $(x,s,y)=(0,\mathbf{1},\mathbf{1})$ and
    iteratively reduces primal infeasibility, dual infeasibility, and the complementarity
    residual $\mu=(s^\top y)/m$. At each iteration, the solver forms the KKT
    system and solves it twice with different right-hand sides, once for the predictor direction and once for the corrector direction.
    If the affine step length is sufficiently large, as controlled by the user-selected corrector threshold, the second solve may be skipped.
    The impact of the corrector threshold is visualized in Figure \ref{fig:corr}. It shows that with a high corrector threshold, the convergence is more variable across different objective directions.
    In the low corrector threshold regime, objective directions which have the same optimal vertex are much more tightly packed. Besides the corrector threshold, \lpviz{} also allows users to control the $\alpha_{\max}$ parameter which scales the final step length. In practice, $\alpha_{\max}$ is set close to one, while in \lpviz{} it defaults to $0.1$ to better visualize the convergence path; otherwise, the problems are solved in too few iterations, limiting the pedagogical value of the visualization.
    Finally, in the 3D view, \lpviz{} places IPM iterates at height $\mu=(s^\top y)/m$, shown in Figure \ref{fig:teaser:3d}.

\begin{figure}[t!]
      \begin{subfigure}[b]{0.48\columnwidth}
        \centering
        \includegraphics[width=\textwidth, alt={Screenshot of lpviz in 3D mode, showing Central Path in trace mode with height set to the barrier parameter mu.}]{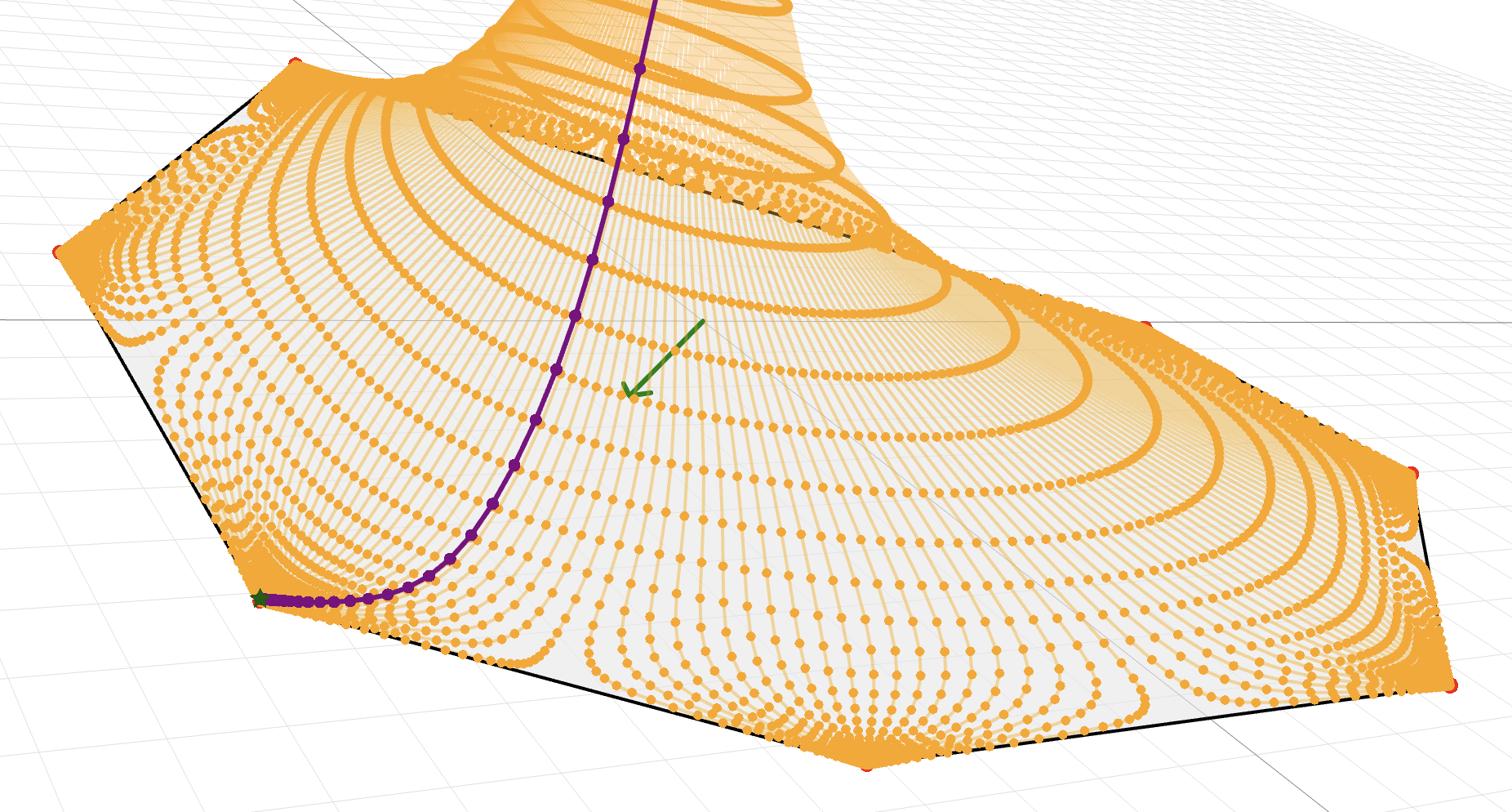}
        \caption{Central Path}
        \label{fig:central3d}
      \end{subfigure}%
      \hfill%
      \begin{subfigure}[b]{0.48\columnwidth}
        \centering
        \includegraphics[width=\textwidth, alt={Screenshot of lpviz in 3D mode, showing a PDHG trajectory with height proportional to the KKT residual epsilon and colors indicating inferred bases.}]{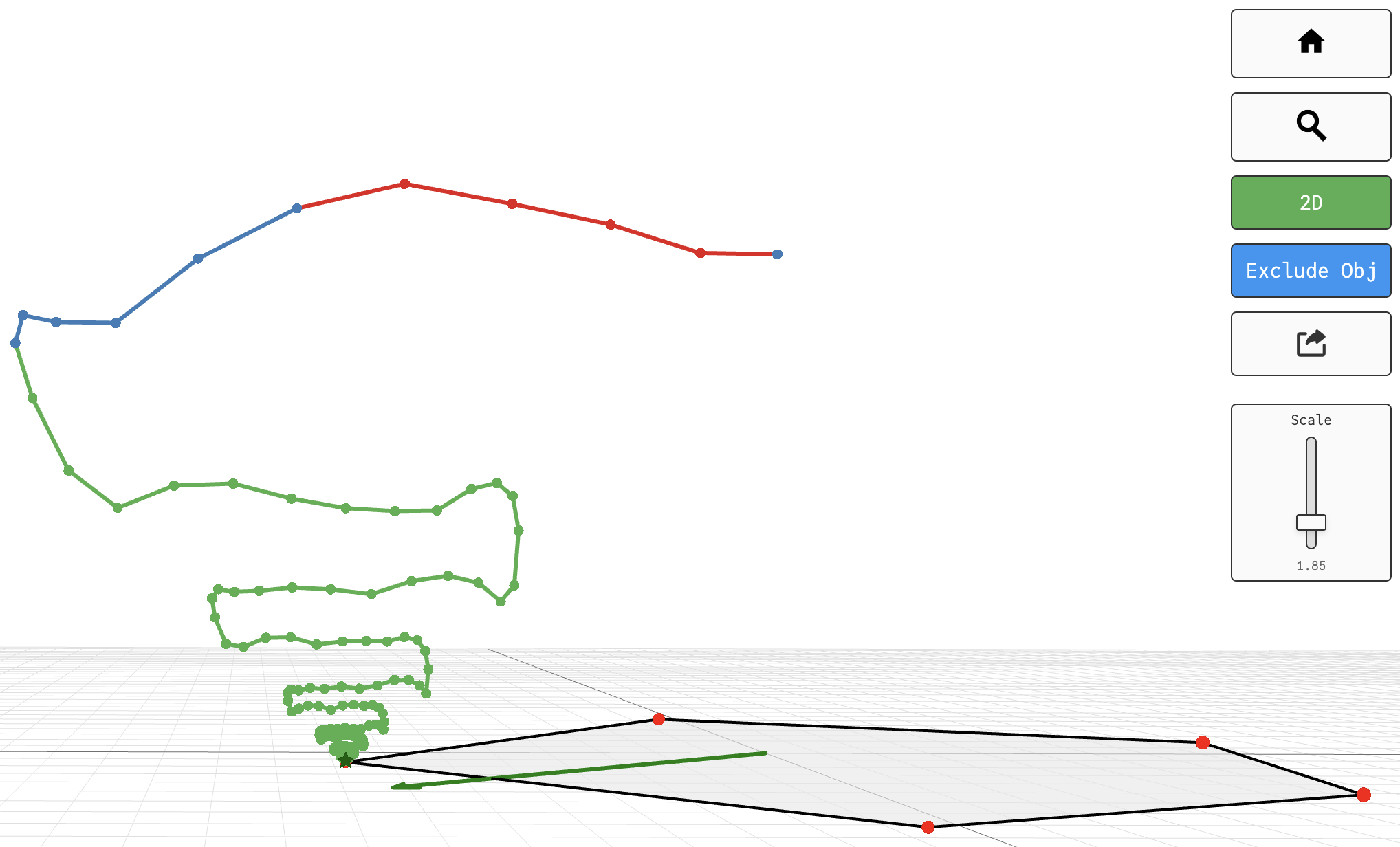}
        \caption{PDHG}
        \label{fig:pdhg3d}
      \end{subfigure}
      \caption{In the 3D view, the Z axis is solver-specific: $\mu$ for Central Path and $\epsilon$ for PDHG.}
      \label{fig:solver3d}
    \end{figure}

\paragraph{Primal-Dual Hybrid Gradient}
First-order methods for LP have become popular recently due to their amenability to GPU acceleration since they require only matrix-vector products and element-wise operations. Although they avoid solving linear systems entirely, making each iteration relatively inexpensive, they typically require many iterations and struggle to converge to high tolerances.
    \lpviz{} implements Chambolle-Pock PDHG \cite{chambolle2011first} in two
    modes. The Equality mode is most common in the literature, while the Inequality mode is a better fit for \lpviz{}. Using Equality mode requires a reformulation that introduces auxiliary variables (similar to the Simplex reformulation), making the trajectories appear more complicated; this is visualized in Figure \ref{fig:eqineq}.
    Both modes use fixed step sizes,
    stop when $\epsilon_k$ (the maximum of a normalized primal residual, dual
    residual, and duality gap) falls below tolerance, and optionally apply
    restarted Halpern iteration \cite{lu2024restarted}. 
    Both modes also support coloring iterates by an inferred basis \cite{liu2026new}. In the
    3D view, \lpviz{} places PDHG iterates at height $\epsilon_k$, as shown in
    Figure \ref{fig:pdhg3d}.

\paragraph{Central Path}
    Although Central Path is not a practical solver for real-world LP, \lpviz{} includes it due to its theoretical and pedagogical value. The Central Path mode plots as iterates the solutions of a sequence of primal log-barrier problems
    \mbox{$
    f_\mu(x)=c^\top x+\mu\sum_i \log(b_i-a_i^\top x),
    $}
    with $\mu$ decreasing over a log-spaced schedule from $10^3$ to $10^{-5}$. Each subproblem is warm-started from the previous solution and solved using
    Newton ascent with a backtracking line search. Finally, in 3D mode Central
    Path iterates are placed at height $\mu$, as shown in Figure \ref{fig:central3d}.
    
\section{Conclusion}
This paper presented \lpviz{}, a browser-based interactive visualization tool
for linear programming algorithms. By combining deep interactivity and reactivity with in-browser execution of solver algorithms, \lpviz{} allows users to explore how different LP algorithms behave in an intuitive and responsive way. A key feature of \lpviz{} is that it brings together four families of LP solver algorithms under one unified interface, making it easy to compare their trajectories and convergence behavior on the same problem. Combined with interactive modification, this allows users to not only compare different solvers, but also develop intuition for the sensitivity of each solver family to changes in problem data and solver settings. Leveraging modern web technologies and best practices, the solver and rendering implementation in \lpviz{} gives users instant feedback without requiring installation steps or expensive backend infrastructure. 
\lpviz{} is open-source, permissively licensed, and freely available on any device with a web browser at \url{https://lpviz.net}.


\acknowledgments{%
	The authors thank Mathieu Tanneau for developing the initial version of the IPM implementation. Additionally, the authors thank Kevin Wu for using \lpviz{} in the Spring 2025 offering of ISYE 3133 Engineering Optimization at the Georgia Institute of Technology.

This material is based upon work supported by the National Science Foundation under Grant No. 2112533 and Grant No. DGE-2039655. Any opinions, findings, and conclusions or recommendations expressed in this material are those of the author(s) and do not necessarily reflect the views of the National Science Foundation.
}

\bibliographystyle{abbrv-doi-hyperref}

\bibliography{template}

\end{document}